\documentclass[twocolumn,prc,aps,floatfix]{revtex4}
\usepackage[dvips]{graphicx}
\begin{document}
\title{Variation of fundamental constants}
\author{V.V. Flambaum}
\affiliation{
 School of Physics, The University of New South Wales, Sydney NSW
2052, Australia, and Argonne National Laboratory, Physics Division,
 9700 S. Cass Av, Argonne, IL 60439, USA}

\begin{abstract}
Theories unifying gravity with other interactions suggest
temporal and spatial variation of the fundamental "constants"
in expanding Universe. The spatial variation can
explain  a fine tuning of the fundamental
constants which allows humans (and any life) to appear.
We appeared in the area of the Universe where the values
 of the fundamental constants are consistent with our existence.
 
 We present a review of recent works devoted to the variation
of the fine structure constant $\alpha$,
strong interaction and fundamental masses.
 There are some hints for the variation in  quasar absorption spectra,
 Big Bang nucleosynthesis, and Oklo natural nuclear reactor data.
 
   A very promising method to search for the variation
of the fundamental constants consists in comparison of
different atomic clocks. Huge enhancement of
the variation effects happens in transition between
accidentally  degenerate  atomic and molecular  energy levels.
A new idea is to build a ``nuclear'' clock based on the ultraviolet
transition between  very low excited state and ground
state in Thorium nucleus. This may allow to improve sensitivity
to the variation  up to 10 orders of magnitude!
                                                                                                    
  Huge enhancement of the variation effects is also possible
in cold  atomic and molecular collisions near Feschbach resonance.

\end{abstract}

\maketitle


\section{Introduction}

 The possible variation of the fundamental constants
of nature is currently a very popular research topic.
 Theories unifying gravity and other interactions suggest
 the possibility of spatial
and temporal variation of physical ``constants'' in the Universe (see,
e.g. \cite{Marciano,Uzan}). Current interest is high because in superstring
theories -- which have additional dimensions compactified on tiny scales --
any variation of the size of the extra dimensions results in changes in the
3-dimensional coupling constants.  At present no mechanism for keeping the
spatial scale static has been found (for example, our three
``large'' spatial dimensions increase in size).  Moreover, there exists a
mechanism for making all coupling constants and masses of elementary
particles both space and time dependent, and influenced by local
circumstances (see e.g. review \cite{Uzan}).
  The variation of coupling constants can be
non-monotonic (for example, damped oscillations).

Recent observations have produced several  hints for the variation of the
fine structure constant, $\alpha=e^2/\hbar c$,  
strength constant of the strong interaction and masses
in Big Bang nucleosynthesis, quasar absorption spectra and Oklo
natural nuclear reactor data
(see e.g.\cite{Murphy,Dmitriev,Lam,Ubach}) .
However, a majority
of publications report only limits on  possible variations
 (see e.g. reviews \cite{Uzan,karshenboim}).
 A very sensitive method to study the
 variation in a laboratory
 consists of the comparison of different optical and microwave atomic clocks
(see  recent measurements in
 \cite{prestage,Marion 2003,Bize 2005,Peik 2004,Bize 2003,
Fischer 2004,Peik 2005}). Huge enhancement of the relative variation effects 
 can be obtained
in transition between the almost degenerate levels in 
 atoms \cite{dzuba1999,Dy,nevsky,budker},
molecules \cite{mol} and nuclei \cite{th}.

\section{Optical spectra}
\subsection{Comparison of quasar absorption spectra with laboratory spectra}
To perform measurements of $\alpha$ variation by comparison of cosmic and
 laboratory optical spectra  we developed a new approach
\cite{dzubaPRL,dzuba1999} which improves the sensitivity to a
variation of $\alpha$ by more than an order of magnitude.
  The relative value of any relativistic
corrections to atomic transition frequencies is proportional to
$\alpha^2$. These corrections can exceed the fine structure interval
between the excited levels by an order of magnitude (for example, an
$s$-wave electron does not have the spin-orbit splitting but it has the
maximal relativistic correction to energy). The relativistic corrections
vary very strongly from atom to atom and can have opposite signs in
different transitions (for example, in $s$-$p$ and $d$-$p$
transitions). Thus, any variation of $\alpha$ could be revealed by
comparing different transitions in different atoms in cosmic and laboratory
spectra.

This method provides an order of magnitude precision gain compared to
measurements of the fine structure interval.  Relativistic many-body
calculations are used to reveal the dependence of atomic frequencies on
$\alpha$ for a range of atomic species observed in quasar absorption
spectra \cite{dzuba1999,dzubaPRL,Dy,q}.  It is convenient to present results for the
transition frequencies as functions of $\alpha^2$ in the form
\begin{equation}
\label{q1}
\omega = \omega_0 + q  x,
\label{omega}
\end{equation}
where $x = (\frac{\alpha}{\alpha_0})^2 - 1 \approx
 \frac{2 \delta \alpha}{\alpha}$ and
 $\omega_0 $ is a laboratory frequency of a particular transition.
We stress that the second  term contributes only if $\alpha$
deviates from the laboratory value $\alpha_0$. 
We performed accurate many-body calculations of the coefficients $q$
for all transtions of astrophysical interest (strong E1 transtions from the
ground state)  in Mg, Mg II, Fe II, Cr II, Ni II, Al II, Al III, Si II,
and Zn II. It is very important that this  set of transtions
contains three large classes : positive shifters (large positive coefficients
$q > 1000 $ cm$^{-1}$), negative shifters (large negative coefficients
$q <- 1000 $ cm$^{-1}$) and anchor lines with small values of $q$.
This gives us an excellent control of systematic errors
since systematic effects do not ``know'' about sign and magnitude
of $q$. Comparison of cosmic frequencies $\omega$  and
 laboratory frequencies $\omega_0$ allows us
to measure $\frac{ \delta \alpha}{\alpha}$.

 Three independent samples of data contaning 143
absorption  systems spread over red shift range $0.2 <z < 4.2$.
 The fit of the data gives \cite{Murphy}
 is $\frac{ \delta \alpha}{\alpha}=
(-0.543 \pm 0.116) \times 10^{-5}$. If one assumes the linear dependence
of $\alpha$ on time, the fit of the data gives $d\ln{\alpha}/dt=
(6.40 \pm 1.35) \times 10^{-16}$ per year
 (over time interval about 12 billion years).
 A very extensive search for possible
systematic errors has shown that known systematic effects can not explain
 the result (It is still  not completely excluded that the effect may be
 imitated by a large change of abundances of isotopes
 during last 10 billion years. 
We have checked that different isotopic abundances for any single
element can not imitate the observed effect. It may be an 
improbable ``conspiracy'' of several elements).

 Recently our method and calculations
 \cite{dzuba1999,dzubaPRL,Dy,q}
were used by two other groups \cite{chand}. However, they have
not detected any variation of $\alpha$. Most probably, the difference
is explained by some undiscovered systematic effects. However,
another explanation is not excluded.
   These results of \cite{Murphy} are based on the data from the
 Keck telescope which is located in the Northen hemisphere  (Hawaii).
 The results of \cite{chand}
are based on the data from the different telescope (VLT) located
in the Southern hemisphere (Chile). Therefore, the difference in the results
may be explained by the spatial variation of $\alpha$. 

    Using opportunity I would like to ask for new, more accurate laboratory
 measurements of  UV transition frequencies which have been observed
in the quasar absorption spectra. The ``shopping list'' is presented
in \cite{shopping}. We also need the laboratory measurements of 
isotopic shifts  - see \cite{shopping}. We have performed very complicated
calculations of these isotopic shifts \cite{isotope}. However, the
 accuracy of  these calculations in atoms and ions with open d-shell
(like Fe II, Ni II, Cr II, Mn II, Ti II) may be very low. The measurements
 for at list few lines are needed to test these calculations.
These measurements would be very important for a study of  evolution of
 isotope abundances in the Universe, to exclude the systematic effects
 in the search for 
$\alpha$ variation and to test models of nuclear reactions in stars
and supernovi.
\subsection{Optical cloks}
Optical clocks also include transitions which have positive, negative
or small constributions of the relativistic corrections to frequencies.
We used the same methods of the relativistic many-body calculations
to  calculate the dependence on $\alpha$ \cite{dzuba1999,clock,Dy}.
 The coefficients
$q$ for optical clock transitions  may be substantially larger than
in cosmic transitions since the clock transitions are often in  heavy atoms
(Hg II, Yb II, Yb III, etc.) while cosmic spectra contain mostly light
atoms lines ($Z <33$). The relativistic effects are proporitional
to $Z^2 \alpha^2$.    

\section{Enhanced effects of $\alpha$ variation in atoms}
An  enhancement of the relative effect of $\alpha$ variation can be obtained
in transition between the almost degenerate levels in Dy atom
 \cite{dzuba1999,Dy}.
These levels move in opposite directions if  $\alpha$ varies. The relative
variation may be presented as $\delta \omega/\omega=K \delta \alpha /\alpha$
 where the coefficient $K$ exceeds $10^8$. Specific
 values of $K=2 q/\omega$ are different for different 
 hyperfine components and isotopes which have different $\omega$;
 $q=30,000$ cm$^{-1}$,  $\omega \sim 10^{-4}$ cm$^{-1}$. 
 An experiment is currently underway to place limits on
$\alpha$ variation using this transition \cite{budker}.
Unfortunately, one of the levels has  quite a large linewidth
and this limits the accuracy.

Several enhanced effects of $\alpha$ variation in atoms have been calculated
in \cite{nevsky}.

\section{Enhanced effects of $\alpha$ variation in molecules}
 The relative effect of $\alpha$ variation 
in microwave transitions between very close and narrow
rotational-hyperfine levels
may be enhanced 2-3 orders of magnitude in diatomic molecules
with unpaired electrons
like LaS, LaO, LuS, LuO, YbF and similar molecular ions \cite{mol}.
 The enhancement is a result
of cancellation between the hyperfine and rotational intervals; 
$\delta \omega/\omega=K \delta \alpha /\alpha$
 where the coefficients $K$ are between 10 and 1000.

\section{Variation  of the strong interaction}

 The hypothetical unification of all interactions implies that a variation
in $\alpha$ should be accompanied by a variation of the strong interaction
strength and the fundamental masses. For example, the grand unification model
discussed in Ref. \cite{Marciano} predicts the quantum chromodynamics
(QCD) scale $\Lambda_{QCD}$ (defined as the position of the Landau pole in
the logarithm for the running strong coupling constant,
$\alpha_s(r) \sim 1/\ln{(\Lambda_{QCD} r/\hbar c)}$) is modified as
 ${\delta\Lambda_{QCD}}/{\Lambda_{QCD}} \approx 34 \;{\delta\alpha}/{\alpha}$.
The variations of quark mass $m_q$ and electron masses $m_e$
 ( related
to  variation of the  Higgs vaccuum field which generates fundamental
 masses)  in this model are given by 
${\delta m}/{m} \sim 70 \; {\delta\alpha}/{\alpha} $, giving an estimate of
 the variation for the dimensionless ratio
\begin{equation}\label{eq:alpha}
\frac{\delta (m/\Lambda_{QCD})}{(m/\Lambda_{QCD})} \sim
 35\frac{\delta\alpha}{\alpha}
\end{equation}
The coefficient here is model dependent but large values are generic
 for grand unification models in which modifications come from high energy
 scales; they appear because the running strong-coupling constant and
 Higgs constants  (related to mass) run faster than $\alpha$. If these models
 are correct, the variation in electron or quark masses and
 the strong interaction scale
 may be easier to detect than a variation in $\alpha$. One can only measure
 the variation of dimensionless quantities.
 The variation of $m_q/\Lambda_{QCD}$
 can be extracted from consideration of Big  Band nucleosynthesis,
 quasar absorption spectra and the Oklo natural nuclear reactor, which was
 active about 1.8 billion years ago \cite{Flambaum Shuryak 2002}.
 There are some
 hints for the variation in Big Bang Nucleosynthesis
 ($\sim 10^{-3}$ \cite{Dmitriev}) and
Oklo ($\sim 10^{-9}$ \cite{Lam}) data.
 The results for the variation have  small statistical errors
(4 standard deviations from zero for BBN and  11 standard deviations for Oklo).
However, it may be hard to prove that there are no other explanations
for the deviations since both phenomena are very complicated. 

 The proton mass is proportional to $\Lambda_{QCD}$
 ($M_p \sim 3 \Lambda_{QCD}$),
therefore, the measurements of the variation of the electron-to-proton
mass ratio $\mu=m_e/M_p$ is equivalent to the measurements of the variation of 
$X_e=m_e/\Lambda_{QCD}$. Two new results have been obtained recently
using quasar absorption spectra. In our  paper \cite{tzana}
 the varition of the
ratio of the hydrogen hyperfine frequency to optical frequencies in ions have
been measured. The result is consistent with no variation of
 $X_e=m_e/\Lambda_{QCD}$. However, in the most recent
paper \cite{Ubach} the variation
 was detected
at the level of 4 standard deviations: $\frac{ \delta X_e}{X_e}=
\frac{ \delta \mu}{\mu}=
(-2.4 \pm 0.6) \times 10^{-5}$. This result is based on the  hydrogen molecule
 spectra. Note, however, that the difference between the zero result
 of \cite{tzana} and non-zero result of \cite{Ubach} may be explained by
 a space-time variation of $X_e$. The variation of $X_e$ in \cite{Ubach}
is substantially larger than the variation of $\alpha$ measured in
 \cite{Murphy, chand}. This may be considered as an argument in favour
of Grand Unification theories of the variation \cite{Marciano}.

\section{Microwave clocks}
 Karshenboim
 \cite{Karshenboim 2000} has pointed out that measurements of ratios
 of hyperfine structure intervals in different atoms are sensitive to
 variations in nuclear magnetic moments. However, the magnetic moments
are not the fundamental parameters and can not be directly compared with
 any theory of the variations. Atomic and nuclear calculations are needed 
for the interpretation of the measurements. We have performed both
atomic calculations of $\alpha$ dependence \cite{dzuba1999,clock,Dy} and
 nuclear calculations of $X_q=m_q/\Lambda_{QCD}$ dependence \cite{tedesco}
 for all microwave transitions of current experimental interest including
hyperfine transitions in $^{133}$Cs, $^{87}$Rb, $^{171}$Yb$^+$,
$^{199}$Hg$^+$, $^{111}$Cd, $^{129}$Xe, $^{139}$La, $^{1}$H,  $^{2}$H and
 $^{3}$He. The  results for the dependence of the transition frequencies
 on variation
of $\alpha$, $X_e=m_e/\Lambda_{QCD}$ and  $X_q=m_q/\Lambda_{QCD}$
 are presented in Ref.\cite{tedesco} (see the final results in the Table IV
 of Ref.\cite{tedesco}). Also, one can find there experimental
 limits on these variations  which follow from the recent measurements. 
The accuracy is approaching $10^{-15}$ per year. This may be compared
to the sensitivity $\sim 10^{-5}-10^{-6}$ per $10^{10}$ years obtained using 
the quasar absorption spectra.

\section{Enhanced effect of  variation of $\alpha$
and strong interaction in UV transition of $^{229}$Th nucleus (nuclear clock)}

A very  narrow level  $(3.5\pm 1)$ eV above the ground state exists
 in $^{229}$Th nucleus \cite{th1}
 (in \cite{th6} the energy is $(5.5\pm 1)$ eV ). The position
of this level was determined from the energy differences of many high-energy
$\gamma$-transitions (between 25 and 320 KeV) to the ground and excited
 states. The subtraction  produces
the large  uncertainty in the position of the 3.5 eV excited state.
 The width of this level is estimated to be
about $10^{-4}$ Hz \cite{th2}. This would explain why it is so hard to find
 the direct radiation in this very weak  transition. The direct
 measurements have
 only given experimental limits on the width and energy  of this transition
 (see e.g. \cite{th3}). A detailed discussion of the measurements
 (including several unconfirmed
 claims of the  detection of the direct radiation ) is presented
 in Ref.\cite{th2}. However, the search for the direct radiation continues
\cite{private}. 

  The  $^{229}$Th transition is very narrow and can be investigated
 with laser spectroscopy.
 This makes $^{229}$Th a possible reference for an
 optical clock of very high accuracy, and opens a new possibility
for a laboratory search for the varitation of the fundamental constants
\cite{th4}.

As it is shown in Ref. \cite{th} there is an additional very important
 advantage.
The relative effects of variation of
 $\alpha$ and $m_{q}/\Lambda_{QCD}$ are enhanced by 5-6 orders of magnitude.
 The  estimate for the relative variation of the $^{229}$Th
 transition frequency  is
 \begin{equation}\label{deltaf}
\frac{\delta \omega}{\omega} \approx 10^5 (
4 \frac{\delta \alpha}{\alpha} +  \frac{\delta X_q}{X_q }
 -10 {\delta X_s \over X_s})\frac{3.5\,eV }{\omega}
\end{equation} 
where $X_q=m_q/\Lambda_{QCD}$, $X_s=m_s/\Lambda_{QCD}$,
$m_q=(m_u+m_d)/2$ and $m_s$ is the strange quark mass.
Therefore, the Th  experiment would
have the potential of improving the  sensitivity to temporal
variation of the fundamental
constants by many orders of magnitude.

Note that there are other narrow low-energy levels in nuclei,
 e.g. 76 eV level in $^{235}U$ with the 26.6 minutes lifetime
 (see e.g.\cite{th4}). One may expect a similar  enhancement there.
Unfortunetely, this level can not be reached with usual lasers. In principle,
 it may be investigated using a free-electron laser or synchrotron radiation.
However, the accuracy
of the frequency measurements is much lower in this case.

\section{Enhancement of variation of
fundamental constants in ultracold atom and molecule systems near
Feshbach resonances}
Scattering length $A$, which can be measured in Bose-Einstein condensate
and Feshbach molecule experiments, is extremely sensitive to the
variation of the
electron-to-proton mass ratio $\mu=m_e/m_p$ or $X_e=m_e/\Lambda_{QCD}$
 \cite{chin}.
\begin{equation}\label{d_a_final}
\frac{\delta A}{A}=K\frac{\delta\mu}{\mu}=K\frac{\delta X_e}{X_e},
\end{equation}
 where $K$ is the  enhancement factor.
For example, for Cs-Cs collisions we obtained
 $K\sim 400$. With the Feshbach resonance, however, one is
given the flexibility to adjust position of the resonance using
external  fields.  Near a narrow magnetic or an optical Feshbach resonance
 the enhancement factor $K$ may be increased by many orders of magnitude. 

This work is supported by the Australian Research Council
 and Department of Energy, Office of Nuclear Physics,
 Contract No. W-31-109-ENG-38.
  

\end{document}